\documentclass[%
reprint,
superscriptaddress,
nofootinbib,
 amsmath,amssymb,
 aps,
 prd,
longbibliography,
]{revtex4-2}
\usepackage{amsmath, amssymb}
\usepackage{listings}
\usepackage{graphicx}
\usepackage{dcolumn}
\usepackage{bm}
\usepackage{float,color}
\usepackage{xcolor}
\usepackage[normalem]{ulem}
\usepackage{tabularx}
\usepackage{mathtools} 
\usepackage{multirow}
\usepackage{subfigure}
\usepackage{hyperref}

\begin{document}

\title{
Impact of unmodeled eccentricity on the tidal deformability measurement and implications for gravitational wave physics inference}
\author{Poulami Dutta Roy} \email{poulami@cmi.ac.in}
\affiliation{Chennai Mathematical Institute, Siruseri 603103, Tamil Nadu, India}
\author{Pankaj Saini}\email{pankajsaini@cmi.ac.in}
\affiliation{Chennai Mathematical Institute, Siruseri 603103, Tamil Nadu, India}
\date{\today}

\begin{abstract}
With the expected large number of binary neutron star (BNS) observations through gravitational waves (GWs), third-generation GW detectors, Cosmic Explorer (CE) and Einstein Telescope (ET), will be able to constrain the tidal deformability, and hence the equation of state (EoS) of neutron star (NS) with exquisite precision. A subset of the detected BNS systems can retain residual eccentricity in the detector frequency band. We study the systematic errors due to unmodeled eccentricity in the tidal deformability measurement and its implications for NS EoS and redshift measurement via the Love siren method. We find that the systematic errors in the tidal deformability parameter exceed the statistical errors at an eccentricity of $\sim 10^{-3}$ ($\sim 3\times 10^{-4}$) at $10$Hz reference GW frequency for CE (ET). We show that these biases on tidal deformability parameter can significantly bias the NS EoS inference. Furthermore, the error on tidal deformability propagates to the source frame NS mass, which in turn biases the redshift inference. For CE, the redshift inference is significantly biased at an eccentricity of $\sim 10^{-3}$ (at a reference frequency of $10$Hz). We also study the implications of biased tidal deformability in testing the Kerr nature of black holes. Systematic error on the tidal deformability parameter leads to a non-zero value of tidal deformability for binary black holes, indicating a false deviation from the Kerr nature. Finally, we show that including eccentricity in the waveform model increases the statistical errors in tidal deformability measurement by a factor of $\lesssim 2$. Our study, therefore, highlights the importance of using accurate eccentric waveform models for GW parameter inference.
\end{abstract}

\maketitle
\section{Introduction}
Third-generation (3G) gravitational wave (GW) detectors Cosmic Explorer (CE)~\cite{LIGOScientific:2016wof, Reitze:2019iox} and Einstein Telescope (ET)~\cite{Punturo:2010zz} will have an order of magnitude better sensitivity compared to Advanced LIGO~\cite{LIGOScientific:2014pky} and Advanced Virgo~\cite{VIRGO:2014yos}. These detectors are expected to detect $\mathcal{O}(10^3\mbox{--}10^5)$ binary neutron star (BNS) and $\mathcal{O}(10^4\mbox{--}10^5)$ binary black hole (BBH) systems per year with high signal-to-noise ratio (SNR)~\cite{PhysRevD.100.064060, Evans:2023euw,Gupta:2023lga}. Due to the rapid decay of eccentricity caused by the emission of gravitational waves (GWs)~\cite{Peters:1964zz}, most binaries are expected to be circularized when observed in ground-based detectors. To the leading order in the small eccentricity limit, the orbital eccentricity decreases with the GW frequency as $e_t/e_0 \approx (f/f_0)^{-19/18}$~\cite{Yunes:2009yz}. Here $e_t$ is the ``time-eccentricity" when binary emits at the dominant (second) mode of the GW frequency $f$ and $e_0$ is the initial value of $e_t$ defined at a reference frequency of $f_0$. (Throughout the paper we define $e_0$ at a reference GW frequency of $f_0=10$Hz.) N-body simulations suggest that a subset of binaries can retain non-negligible eccentricity $(e_0>0.1)$, depending on their formation history, while entering the frequency band of ground-based detectors. 

Various mechanisms have been proposed that can lead to the formation of highly eccentric binaries. Dynamically formed binaries in dense stellar environments such as globular clusters, nuclear star clusters, and near the supermassive black holes can have non-negligible eccentricity at $10$Hz GW frequency~\cite{Wen:2002km, OLeary:2008myb, Antonini:2012ad,  Bae:2013fna, Antonini:2013tea, Antonini:2015zsa, PhysRevLett.120.151101, Zevin:2017evb, PhysRevD.97.103014, Gondan:2017wzd, 2018PhRvD..98l3005R,Rasskazov:2019gjw, Zevin:2018kzq, Zevin:2021rtf, 2022ApJ...925..178H, 2023arXiv230307421D,2024arXiv240216948F}. Around $5\%$ to $10\%$ of all dynamical mergers in globular clusters can give rise to BBHs with initial eccentricity $e_0 \gtrsim 0.1$~\cite{Samsing:2017xmd,PhysRevD.97.103014,PhysRevLett.120.151101}. The rate of dynamically formed BNSs is expected to be relatively low~\cite{Bae:2013fna} (see however \cite{Barr:2024wwl}). Since black holes (BHs) move closer to the center of the cluster due to dynamical friction, they prevent NSs from sinking towards the cluster core, thereby reducing NS's dynamic interactions with other bodies. The observations of BNS in the Milky Way suggest that the BNS are likely to have small eccentricity [$\mathcal{O}(10^{-5})$] at $10$Hz GW frequency~\cite{Lorimer:2008se}. The reanalysis of detected BNS signals GW170817~\cite{LIGOScientific:2017vwq} and GW190425~\cite{LIGOScientific:2020aai} with eccentric waveforms suggests that these systems have $e_0 \leq 0.024$ and $e_0 \leq 0.048$, respectively~\cite{Lenon:2020oza}. Nevertheless, there are proposed mechanisms for the formation of eccentric NS binaries. These include tidal capture and collisions of NSs in globular clusters~\cite{2010ApJ...720..953L}, Zeipel-Lidov-Kozai mechanism \cite{Zeipel:1910,Lidov:1962,Kozai:1962} for a hierarchical triple system~\cite{2011ApJ...741...82T,Seto:2013wwa}, dynamical interaction of NSs in globular clusters~\cite{Grindlay:2005ym,2009A&A...498..329G,East:2012xq,Andrews:2019vou,Ye:2019xvf}, NS-BH mergers in globular cluster~\cite{2013MNRAS.428.3618C}, BNS mergers in young star clusters~\cite{Ziosi:2014sra}, BNS and NS-BH mergers in galactic-nuclei~\cite{2017ApJ...846..146P} and triple star systems in the field~\cite{Hamers:2019oeq}. Population synthesis models predict that up to $4\%$ of the detected BNS in the 3G era will have eccentricity $\approx0.01$ when entering the frequency band of the detectors~\cite{2011A&A...527A..70K}.  

While LIGO can measure $e_0\sim 0.05$ for GW150914-like systems~\cite{Lower:2018seu, Favata:2021vhw}, 3G detectors --- with better low-frequency sensitivity and hence a much longer inspiral in the band --- will be able to measure $e_0\gtrsim 10^{-3}$ for BBHs~\cite{Saini:2023wdk}. BNS systems will spend an even larger number of GW cycles in the detectors' frequency band, allowing us to measure even smaller eccentricities than $e_0\sim10^{-3}$ with exquisite precision.  Current template-based GW searches by LIGO-Virgo-KAGRA do not account for eccentricity. However, various alternate search methods have been proposed to detect eccentric BBHs and BNSs~\cite{Huerta:2013qb, PhysRevD.93.043007, LIGOScientific:2019dag, Ramos-Buades:2020eju, PhysRevD.104.104016, PhysRevD.104.063011, 2023arXiv230703736P}.  

The analysis of an eccentric GW signal using a quasi-circular waveform model will introduce systematic bias in the estimated parameters~\cite{Favata:2013rwa, Favata:2021vhw, Cho:2022cdy, Divyajyoti:2023rht}. As a BNS typically spends a large number of GW cycles while sweeping through the frequency band of the detector compared to BBHs, systematic errors may accumulate over large number of GW cycles and can become dominant even at a very small eccentricity. Reference~\cite{Favata:2013rwa} studied the effect of neglecting eccentricity for BNSs in LIGO and 3G detectors and found that the systematic bias on binary parameters exceeds the statistical errors even for small eccentricity $e_0\sim 10^{-3}\mbox{--} 10^{-2}$. In the LIGO band, an eccentricity of $e_0\gtrsim 10^{-2}\mbox{--}0.1$ can significantly bias the estimated masses and spins of BHs~\cite{Favata:2021vhw, Divyajyoti:2023rht} and NSs~\cite{Cho:2022cdy}. Eccentricity-induced systematic bias can also bias the tests of general relativity (GR) and appear as a false violation of GR~\cite{Saini:2022igm, Bhat:2022amc, Narayan:2023vhm,Shaikh:2024wyn}. These biases become even more severe when performing tests of GR from a catalog of events~\cite{Moore:2021eok, Saini:2023rto}. 

The systematic errors are independent of SNR, while the statistical errors scale as the inverse of SNR. Hence, for loud events (high SNR), the systematic errors can easily become greater than the statistical errors. Since CE and ET will observe GW events with higher SNR, leading to a precise measurement of binary parameters like the tidal deformability, and hence the equation of state of NS~\cite{PhysRevD.97.024049, PhysRevD.105.123032,PhysRevD.106.084056} and the cosmological parameters~\cite{Chatterjee:2021xrm,Dhani:2022ulg,PhysRevD.106.123529}, therefore, systematic errors can easily bias these  parameters.  

{\it In this work, we study the impact of unmodeled eccentricity on the measurement of the tidal deformability parameter in 3G detectors. In particular, we study the implications of the biased estimation of tidal deformability on the NS EoS, redshift measurement through the Love siren technique, and the test of the Kerr nature of black holes.} Since eccentricity and tidal deformability are relatively low-frequency and high-frequency effects, respectively, their measurement is naively expected to have negligible effects on each other. However, we show that the precise measurement of the tidal deformability in 3G detectors can be biased even for a very small (unmodeled) eccentricity.
\begin{figure*}[t]
    \centering
    \begin{subfigure}[]
  {\includegraphics[width=0.48\textwidth]
  {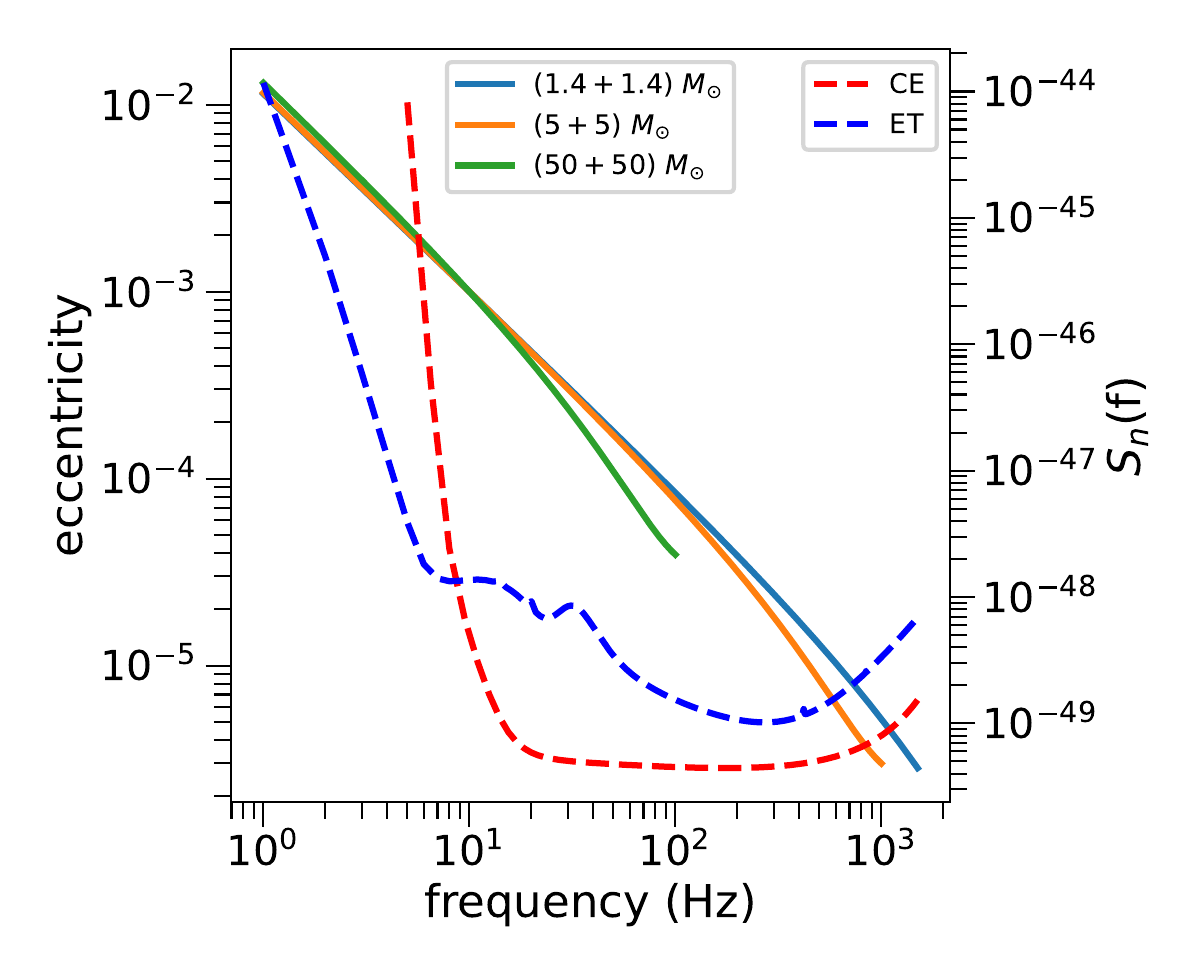}\label{fig:ecc_sensitivity}}
  \end{subfigure}
  \begin{subfigure}[]
 {\includegraphics[width=0.48\textwidth]{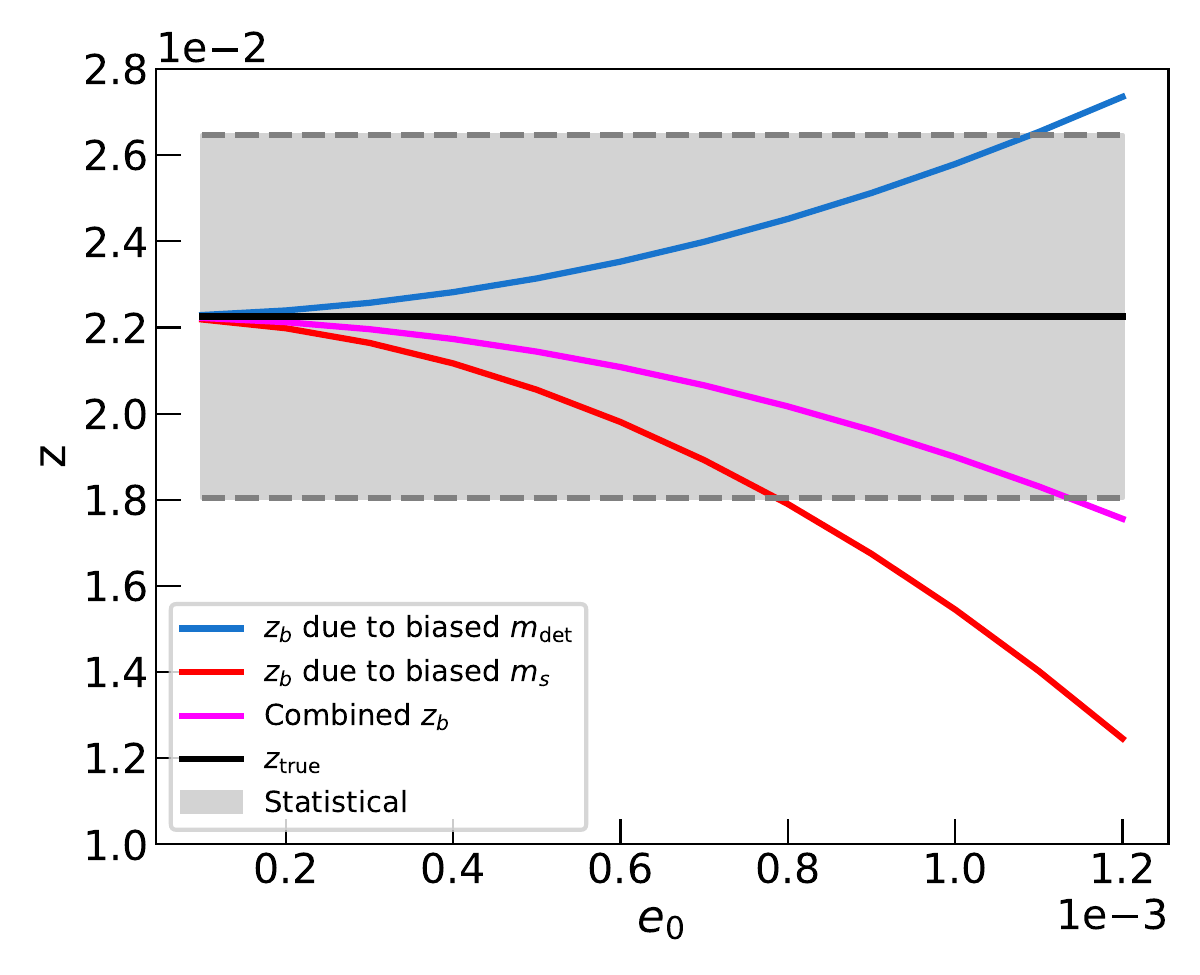}\label{fig:z_bias}}
  \end{subfigure}
  \caption{(Color online) (a) Solid lines show the decay of eccentricity following Eq.~\eqref{eq:ecc_decay} as a function of frequency for different mass binaries. Dashed lines represent the sensitivity curves for CE and ET. BNS system retains a higher value of eccentricity at a given frequency compared to BBHs. (b) The deviation of biased redshift $(z_b)$ (due to neglecting eccentricity) from its true value as a function of initial eccentricity $e_0$ (defined at $10$Hz) for a BNS in CE. We consider an equal mass BNS $(1.4+1.4)M_{\odot}$ with dimensionless spins magnitude $\chi_{1,2}=0.05$ located at $100$ Mpc. The shaded region indicates the $1\sigma$ statistical width around the true value of redshift. The systematic error in $\Tilde{\Lambda}$ biases the source frame NS mass (for a given EoS), which propagates to redshift inference through the Love siren technique. The pink curve shows the combined $z_b$ due to biased source frame mass $m_s$ and detector frame mass $m_{\rm det}$. The plot is discussed in detail in Sec.~\ref{sec:Love siren}.}
\end{figure*}
\subsection{Implications of biased tidal deformability}
\subsubsection{Inference of NS EoS}
During the later stages of NS binary inspiral, the tidal interactions resulting from their finite size effect start coming into play. These effects are imprinted in the emitted GW signal and carry information about the internal structure of the stars \cite{PhysRevD.99.083016,PhysRevD.101.124014}. The corrections due to tidal deformations appear at fifth post-Newtonian (PN) order.\footnote{In a PN series, a term proportional to $v^{2 n}$ relative to the {\it Newtonian term} (proportional to $v^{-5}$) is called the $n$PN order term.} As the name suggests, tidal deformability measures the quadrupolar deformation of a star in response to the external field exerted by its companion in the case of binary. It acts as a proportionality constant relating the object's quadrupole moment with the external tidal field~\cite{Kokkotas:1995, Flanagan:2007ix,Hinderer:2007mb,PhysRevD.81.123016}. 

The observation of GWs from NS binaries can provide valuable information about the nuclear EoS~\cite{Flanagan:2007ix}. Observations of GW170817 and GW190425 have already put constraints on the tidal deformability, which provided information about the NS EoS~\cite{LIGOScientific:2018cki,Biswas:2021paf}. Additionally, Pulsar observations provide information for the NS EoS by precisely measuring the mass and the radius of NS~\cite{Riley:2019yda, Riley:2021pdl,Miller:2021qha}. The measurement of macroscopic properties such as mass, radius, and tidal deformability allows us to constrain the underlying EoS of NS, which carries information about the microscopic properties of NS matter. In the 3G era, the observations of BNS will tightly constrain NS EoS~\cite{Hinderer:2009ca,Steiner:2010fz, Ozel:2010fw,Lackey:2013axa,PhysRevD.108.122006}. Therefore, a biased measurement of the tidal deformability may lead to a biased inference of the EoS. 

\subsubsection{Measurement of redshift through `Love siren'}
The observation of GWs allows for the direct measurement of the luminosity distance of the source~\cite{Schutz:1986gp, Sathyaprakash:2009xs}. Since GWs do not require any other calibration for distance measurement, these are called `standard candles'. Estimating the Hubble constant $(H_0)$ requires the knowledge of the luminosity distance and the redshift to the source. GWs do not directly provide the redshift information, which must be `inferred' through other channels. Once redshift to the source is known, the measurement of luminosity distance from GWs can be used for estimating cosmological parameters~\cite{Holz_2005,PhysRevD.77.043512,Nissanke_2010,PhysRevD.86.043011,Chen:2017rfc}. 

Various methods have been proposed to measure the redshift to the source. The mergers of BNSs are likely to be associated with electromagnetic (EM) emissions like short-gamma ray bursts and afterglow emissions similar to GW170817~\cite{LIGOScientific:2017vwq}. The electromagnetic observations from these mergers can be used to locate the host galaxy enabling the redshift measurement and hence $H_0$~\cite{Nakar:2005bs, Sathyaprakash:2009xt,Nissanke:2009kt, Nissanke:2013fka, LIGOScientific:2017vwq}. This method is called the `bright siren' method. However, most BBH mergers are not likely to be associated with EM counterparts, rendering the redshift measurement via the bright siren method. In the absence of an electromagnetic counterpart, the sky-localization associated with a GW event can be used to identify the potential host galaxies using galaxy surveys. Statistical methods can then be implemented to combine the $H_0$ measurement from the potential host galaxies~\cite{LIGOScientific:2017adf,Borhanian:2020vyr,Gray:2021sew,Gair:2022zsa,PhysRevD.108.043543,Yun:2023ygz}. This is called the `dark siren' technique. Another way to obtain the redshift of the source is to use the features in the source mass distribution of NSs~\cite{Taylor:2011fs,Taylor:2012db} and BHs~\cite {Ezquiaga:2022zkx}. The redshift, in the `spectral siren' method, is inferred by measuring the detector frame mass and jointly fitting the source mass distribution and cosmological parameters~\cite{Mastrogiovanni:2021wsd,Mastrogiovanni:2023emh}. 

In this paper, we focus on yet another technique of redshift inference called the `Love siren' method~\cite{Messenger:2011gi}. This method exploits the fact that the tidal effects are completely described by the source frame NS mass for a given EoS. The detector frame mass is directly measured from the GW data. The redshift of the source is inferred using the relation between the source frame mass and the detector frame mass. This method allows GW observations in 3G detectors alone to constrain cosmological parameters like $H_0$ with sub- percent accuracy~\cite{DelPozzo:2015bna}. A systematic bias on the tidal deformability parameter, due to neglecting eccentricity, will propagate to the source frame mass, leading to a biased inference of redshift.

\subsubsection{Test of Kerr nature of binary black holes}
Lastly, we study the impact of biased tidal deformability measurement on the Kerr nature of the BHs. Within GR, black holes (both static and rotating) have zero tidal deformability~\cite{Damour:2009vw, Binnington:2009bb, Chakrabarti:2013lua,Pani:2015hfa,Landry:2015zfa,PhysRevLett.114.151102,LeTiec:2020spy,Saini:2023gaw} (see however \cite{Chakravarti:2018vlt,DeLuca:2023mio}). On the other hand, exotic compact objects are known to have non-zero tidal deformability~\cite{Uchikata:2016qku,Cardoso:2017cfl,Mendes:2016vdr}. The tidal deformability parameter is directly proportional to Love number which has been widely explored for BHs \cite{PhysRevD.80.084018,PhysRevD.80.044017,PhysRevD.89.124011,PhysRevD.92.024010, PhysRevLett.126.131102,PhysRevD.103.084021} and through the {\em I-Love-Q} relations for NS \cite{Yagi:2013bca,PhysRevD.88.023009,PhysRevD.88.023007,PhysRevLett.112.121101,PhysRevLett.121.091102}. In the context of BHs, our study acts as a null test for BBH nature, with systematic biases producing non-zero tidal deformability leading to false inference of a source as non-BBH. 

\subsection{Eccentricity evolution as a function of GW frequency}
In post-Newtonian theory~\cite{Blanchet:2002av}, the decay of eccentricity with the GW frequency is described by an analytical expression of the form (Eq.(4.17a) in~\cite{Moore:2016qxz}),
\begin{equation}
     e_{t} = e_{0} \bigg(\frac{f}{f_0} \bigg)^{-19/18} \frac{\zeta(M,\eta,f)}{\zeta(M,\eta,f_0)} \,.
     \label{eq:ecc_decay}
\end{equation}
The term $\zeta$ has the PN structure
\begin{equation}
    \zeta = 1+\bigg(-\frac{2833}{2016} + \frac{197}{72} \eta \bigg) v^{2} + \cdots \mathcal{O}(v^6) \,,
\end{equation}
where $v=(\pi M f)^{1/3}$ is the orbital velocity parameter, $M=m_1+m_2$ is the total mass of the source, and $\eta=\frac{m_1m_2}{(m_1+m_2)^2}$ is the symmetric mass ratio. The full expression for $\zeta$ can be found in Ref.~\cite{Moore:2016qxz}. Figure~\ref{fig:ecc_sensitivity} shows the decay of eccentricity as a function of GW frequency in the CE and ET frequency bands. Three distinct lines in different colors represent different mass binaries. If the three systems have the same initial eccentricity, the BNS system retains a higher value of eccentricity (at a particular frequency). Moreover, the BNS system merges at a higher frequency and spends more time in the detector band. The main factors that play a role in the accumulation of systematic error for a given binary are (i) the rate of decay of eccentricity in the frequency band of the detector, (ii) the time of onset of tidal effects, and (iii) the number of GW cycles in the detector band. Though the tidal effects come into play when the stars come sufficiently close to each other, the BNS system can retain a higher value of eccentricity and spend more time in the detector band. Therefore, neglecting eccentricity can potentially bias the tidal deformability parameter which is indeed found to be true in our study.

Figure~\ref{fig:z_bias} shows one of the main results of the paper. It shows the inferred redshift via the Love siren method can be significantly biased due to unmodeled eccentricity. Since the source frame NS mass is obtained from the tidal deformability, the systematic bias on tidal deformability propagates to the source frame NS mass. The plot shows the systematic errors due to source frame and detector frame NS mass as a function of initial eccentricity $e_0$ (defined at a reference frequency of $10$Hz). The redshift bias shows an opposite trend due to source frame and detector frame mass. The combined effect of systematic errors is negative and indicates redshift being underestimated when ignoring eccentricity. The shaded region shows $1\sigma$ statistical width around the true value of redshift. The combined systematic error crosses the statistical errors at $e_0\sim 10^{-3}$. The plot is discussed in further detail in Sec.~\ref{sec:Love siren}.

The rest of the paper is organized as follows. In Sec.~\ref{sec:waveform model}, we describe the frequency-domain waveform model used in the study. Section~\ref{sec:error analysis} explains the Fisher information matrix framework to calculate the statistical errors and Cutler--Vallisneri formalism~\cite{Cutler:2007mi} to calculate the systematic errors. Section~\ref{sec:results} discusses the implications of biased tidal deformability for NS EoS and redshift measurement through `Love siren' technique. In Sec.~\ref{sec:BBH nature}, we show how the biased tidal deformability can lead to a false deviation of the Kerr nature of BBHs. The conclusion of the study is described in Sec.~\ref{sec:conclusion}. Throughout the paper we use $G=c=1$.

\section{Waveform model}\label{sec:waveform model}
The frequency-domain GW strain in stationary phase approximation (SPA)~\cite{Droz:1999qx} can be written as 
\begin{align}
    \Tilde{h}(f) = \mathcal{A}(f) \, e^{i \Psi (f)} = \hat{\mathcal{A}} f^{-7/6} e^{i \Psi(f)} \,,
    \label{eq:waveform}
\end{align}
where 
\begin{equation}
    \hat{\mathcal{A}} = \frac{1}{\sqrt{30} \pi^{2/3}} \frac{\mathcal{M}^{5/6}}{d_L} \,.
\end{equation}
Here $\mathcal{M} = (m_1 m_2)^{3/5}/M^{1/5}$ is the chirp mass, $M=m_1+m_2$ is the total mass of the source, $m_1$ and $m_2$ are the masses of primary and secondary component of the binary, and $d_L$ is the luminosity distance to the source. Note that $\mathcal{M}$ and $M$ are the {\it detector frame} chirp and the total mass of the binary. The {\it detector frame} chirp and total mass are related to the {\it source frame} chirp mass ($\mathcal{M_{\rm s}}$) and total mass ($M_{\rm s}$) as
\begin{equation}
    \mathcal{M} = (1+z)\mathcal{M}_{\rm s}\,, \;\; M = (1+z) M_{\rm s} \,,
\end{equation}
where $z$ is the redshift to the source. Considering flat Lambda-CDM cosmology, $d_L$ and $z$ are related by
\begin{equation}
d_{L} (z) = \frac{(1+z) }{H_0} \int_0^z \frac{dz'}{\sqrt{\Omega_M (1+z')^3 + \Omega_\Lambda}} \,,
\label{eq:redshift}
\end{equation}
where the cosmological parameters are $\Omega_{M} = 0.3065$, $\Omega_\Lambda = 0.6935$ and $h = 0.6790$ with $H_0 =100h$ (km/s)/Mpc~\cite{2020A&A...641A...6P}.

In PN approximation, which is valid in weak-field and small-velocity regime~\cite{Blanchet:2002av}, the SPA phase $\Psi(f)$ in Eq.~\eqref{eq:waveform} can be expanded as a series in powers of orbital velocity parameter $v$ 
\begin{eqnarray}
    \Psi (f)& =& \phi_c + 2 \, \pi \,f \, t_c  +  \frac{3}{128 \, \eta \,v^5} \Big( 1 + \Delta \Psi_{4.5 \mathrm{PN}}^{\mathrm{pp,circ.}} + \nonumber\\ && \Delta \Psi_{4 \mathrm{PN}}^{\mathrm{spin,circ.}} 
    + \Delta \Psi_{3 \mathrm{PN}}^{\mathrm{ecc.}} +\Delta \Psi_{6 \mathrm{PN}}^{\mathrm{tidal}}\Big) \,.
    \label{eq:PNphase}
\end{eqnarray}
Recall that $\eta=\frac{m_1m_2}{(m_1+m_2)^2}$ is the symmetric mass ratio and $v = (\pi M f)^{1/3}$. The term $t_c$ is a kinematical quantity closely related to the time of arrival of the signal at the detector and $\phi_c$ denotes a constant phase. The term $\Delta \Psi_{4.5 \mathrm{PN}}^{\mathrm{pp,circ.}}$ represents the circular point particle contribution to the phase and was recently extended till 4.5PN in~\cite{Blanchet:2023bwj}. We incorporate terms up to 4.5PN order in our analysis which can be written as
\begin{equation}
\Delta \Psi^{\mathrm{pp, circ.}}_{4.5 \mathrm{PN}}(f)= \sum_{k=0}^{9}\left(\phi_k\,v^k+ \right.
    \phi_{kl} v^k \ln v  + 
    \left.\phi_{kl2} v^k \ln^2 v\right) \,,
   \label{eq:PNphase_circ}
\end{equation}
where $\phi_k$ are the PN coefficients which are the functions of source properties such as masses, spins, and tidal deformability. $\phi_{kl}$ and $\phi_{kl2}$ are the PN coefficients which appear with the $\ln$ and $\ln^2$ terms with $\phi_{kl}$ being non-zero only for $k=5, 6, 8,9$ and $\phi_{kl2}$ for $k=8$. The exact expressions for the PN coefficients can be found in~\cite{Buonanno:2009zt,Blanchet:2023bwj}. The spin contribution to the circular part $\Delta \Psi_{4 \mathrm{PN}}^{\mathrm{spin,circ.}}$ can be found in~\cite{Arun:2004hn, Arun:2008kb,Buonanno:2009zt,Mishra:2016whh}. The eccentricity corrections up to 3PN order $\Delta \Psi_{3 \mathrm{PN}}^{\mathrm{ecc.}}$ are taken from the {\tt TaylorF2Ecc} waveform model given in Eq.(6.26) of ~\cite{Moore:2016qxz} and can be written as
\begin{align}
   \Delta \Psi_{3 \mathrm{PN}}^{\mathrm{ecc.}} &= - \frac{2355}  {1462} e_0^2 \left(\frac{v_0}{v}\right)^{19/3} \Big[1+ \Big(\frac{299076223}{81976608} \nonumber \\ 
   &+ \frac{18766963}{2927736}\eta \Big) v^2 + \Big(\frac{2833}{1008} - \frac{197}{36} \eta \Big) v_0^2 \nonumber \\
   & - \frac{2819123}{282600} \pi v^3 + \frac{377}{72} \pi v_0^3 + \cdots + \mathcal{O}(v^6) \Big] ,
\end{align}
where $e_0$ is the value of eccentricity at a reference frequency $f_0$ and $v_0=(\pi M f_0)^{1/3}$. The waveform incorporates leading-order eccentricity corrections  $[\sim \mathcal{O}(e_0^2)]$ in the GW phasing and is valid for eccentricities $\lesssim 0.2$. The waveform model does not account for the eccentric corrections to the GW amplitude. Since GW detectors are much more sensitive to the GW phase than the amplitude, small eccentricity corrections to the amplitude will be less important than the GW phase. The waveform model accounts for the dominant second harmonic of the GW signal and ignores the eccentricity-induced higher harmonics that are expected to be subdominant for small eccentricities. The spins of the two black holes are assumed to be aligned with the orbital angular momentum of the binary (non-precessing).

The leading-order tidal contribution appears at 5PN order \cite{Flanagan:2007ix,Hinderer:2007mb,Vines:2011ud,Favata:2013rwa} and is of the form 
\begin{eqnarray}
    \Delta \Psi_{6 \mathrm{PN}}^{\mathrm{tidal}} \, =  - \frac{39}{2}\Tilde{\Lambda} v^{10} + v^{12} \Big( \frac{6595}{364} \delta\Tilde{\Lambda} - \frac{3115}{64}\Tilde{\Lambda} \Big) \label{eq:PNphase_tidal} \,,
\end{eqnarray}
where $\Tilde{\Lambda}$ is the reduced dimensionless tidal deformability parameter. The parameter $\Tilde{\Lambda}$ can be written in terms of the individual dimensionless tidal deformability parameters $\hat{\lambda}_1, \hat{\lambda}_2$ 
\begin{eqnarray}
    \Tilde{\Lambda} &=& \frac{8}{13} \Big[ (1+ 7 \eta -31 \eta^2) (\hat{\lambda}_1 + \hat{\lambda}_2) \nonumber \\
    && - \sqrt{1-4 \eta} (1+ 9\eta -11 \eta^2)(\hat{\lambda}_1 - \hat{\lambda}_2) \Big] \,.
    \label{eq:tidal_deformability}
\end{eqnarray}
We consider equal mass binaries with $\hat{\lambda}_1 = \hat{\lambda}_2 = \hat{\lambda}$ i.e., $\Tilde{\Lambda} \to \hat{\lambda}$ and $\delta\Tilde{\Lambda} \to 0$. This leads to the reduction in the number of parameters to be estimated~\cite{Favata:2013rwa}. The contribution of $\delta\tilde{\Lambda}$ (compared to the $\tilde{\Lambda}$) is small and can be neglected. Although the tidal contribution to phase has been calculated to even higher PN orders~\cite{Henry:2020ski}, Eq.~\eqref{eq:PNphase_tidal} is sufficient for the present context as the aim is to compute the systematic error due to eccentricity on tidal deformability given that the eccentricity is a low-frequency effect.

\section{Error analysis}\label{sec:error analysis}
In addition to the statistical errors due to detector noise, mismodeling in the waveform model will introduce systematic errors in the estimated parameters. If systematic errors are within the statistical uncertainty, it might be safe to ignore these errors. However, if systematic errors exceed statistical errors, it becomes crucial to understand the extent to which these errors can impact the parameter inference.

We use the {\it Fisher information matrix} framework to calculate the statistical errors on binary parameters~\cite{Finn:1992wt,Cutler:1994ys,Poisson:1995ef}. The Fisher matrix framework is an approximation that is valid in the high SNR limit. In our study, we ensure that we are in a high-SNR regime where Fisher matrix estimates are reasonably accurate. For stationary, Gaussian noise, and in the large SNR limit, the probability distribution of the waveform parameters $\bm \theta$, given the data $d(t)$, can be approximated as
\begin{equation}\label{probability}
 p({\bm \theta}|d) \propto p^{0}({\bm \theta}) \exp\left[ -\frac{1}{2} \Gamma_{ab} (\theta_{a} - \hat{\theta}_{a}) (\theta_{b} - \hat{\theta}_{b}) \right]\,, \end{equation}
where $p^{0}(\bm{\theta})$ is the prior probability of the parameters ${\bm {\theta}}$. The $\hat{\theta}_{a}$ are the ``best-fit" values that maximize the Gaussian likelihood. In the absence of any bias, $\hat{\theta}_{a}$ represents the “true'' values of the source parameters. The Fisher matrix $\Gamma_{ab}$ is defined as
\begin{equation}
    \Gamma_{ab}=2\int_{f_{\rm low}}^{f_{\rm up}}\,\frac{{\tilde h}_{,a}{\tilde h}^{*}_{,b}+{\tilde h}_{,b}{\tilde h}^{*}_{,a}}{S_n(f)} df \,,
\end{equation}
where $*$ denotes complex conjugation, `$,$' denotes the partial derivative with respect to parameter $\theta_a$, $\tilde{h}(f)$ is the Fourier transform of $h(t)$ and $S_n(f)$ is the noise power spectral density (PSD) of the detector. The SNR $\rho$ for a signal $h(t)$ is defined as
\begin{eqnarray}
    \rho^2 = 4 \int_{f_{\rm low}}^{f_{\rm up}} \frac{|\tilde{h}(f)|^2}{S_n (f)} df \,,
\end{eqnarray}
where $f_{\rm low}$ and $f_{\rm up}$ are the lower and upper cut-off frequencies that depend on the detector sensitivity and properties of the source. The covariance matrix $\Sigma_{ab}$ is obtained by taking the inverse of the Fisher matrix and $1\sigma$ statistical error for each parameter ($\sigma_a$) is given by the square root of the diagonal components of the covariance matrix
\begin{eqnarray}\label{eq:covar}
    \Sigma_{ab} = \Gamma^{-1}_{ab}, \,\,\,\, \sigma_a = \sqrt{\Sigma_{aa}} \,.
\end{eqnarray}
The parameters of the quasi-circular waveform model for aligned spins are $\theta_a = \big( t_c, \, \phi_c, \, \ln\mathcal{M}, \, \ln\eta, \, \chi_1, \chi_2,\Tilde{\Lambda}\big)$. $\chi_{1,2}$ are the dimensionless spin magnitudes along the orbital angular momentum.
Constraining the luminosity distance is not a point of concern in this work. The parameter $\hat{\mathcal{A}}$ can be taken as a single parameter, which if included in the parameter space, will decouple from the rest of the Fisher matrix~\cite{Clifford_Will}. Hence $\hat{\mathcal{A}}$ is excluded from the parameter set. We impose Gaussian priors which are added to the diagonal terms of the Fisher matrix as $1/(\delta\theta_a)^2$, $\delta \phi_c = \pi,\, \delta \chi_1 =\delta \chi_2 = 1, \, \delta\Tilde{\Lambda} = 5000$.

The systematic error is defined as the difference between the ``true" value of the parameter $\theta_a^{\rm T}$ and the ``best-fit" value of the parameter $\hat{\theta}_a$ (the peak of the recovered Gaussian probability distribution)
\begin{equation}
    \Delta \theta_a = \theta_a^{\rm T} - \hat{\theta}_a\,,
\end{equation}
which can be computed using the Cutler-Vallisneri formalism \cite{Cutler:2007mi}. The formalism assumes a {\it true} waveform model which describes the true physical system and an {\it approximate} waveform model which is used to model the system. If the approximate waveform $\Tilde{h}_{\rm AP}$ is represented by the approximate amplitude $\mathcal{A}_{\rm AP}$ and approximate phase $\Psi_{\rm AP}$
\begin{equation}
     \Tilde{h}_{\rm AP} = \mathcal{A}_{\rm AP} e^{i \Psi_{\rm AP}} \,, 
     \end{equation}
and the true waveform $\Tilde{h}_{\rm T}$ differs from $\Tilde{h}_{\rm AP}$ in amplitude and phase by $\Delta \mathcal{A}$ and $\Delta\Psi$, respectively, 
\begin{eqnarray}
    \Tilde{h}_{\rm T} = [\mathcal{A}_{\rm AP}+ \Delta \mathcal{A}] e^{i [\Psi_{\rm AP} + \Delta \Psi]} \,,
\end{eqnarray}
then the systematic error can be approximated as~\cite{Favata:2021vhw, Saini:2022igm}
\begin{equation}
    \Delta \theta^a \approx \Sigma^{ab} \Big[(\Delta \mathcal{A} + i\mathcal{A}_{\rm AP} \Delta \Psi) e^{i \Psi_{\rm AP}} \Big\vert \partial_b \Tilde{h}_{\rm AP} \Big] \,,
\end{equation}
where $\Sigma_{ab}$ is the covariance matrix which is calculated using the approximate waveform. In our work, $\tilde{h}_{\rm AP}$ represents the quasi-circular waveform model and $\Delta\Psi = \Delta\Psi_{\rm ecc.}$. We do not account for the eccentricity corrections to the amplitude i.e., $\Delta \mathcal{A} =0$ which are expected to be small for small eccentricities.

We also study the effect of measuring eccentricity along with other binary parameters. In particular, we examine the effect of including eccentricity in the waveform model on the measurement of the tidal deformability parameter $\Tilde{\Lambda}$. For this, we include $e_0$ in the parameter space $ \theta_a = ( t_c, \, \phi_c, \, \ln\mathcal{M},\ln\eta, \, \chi_1, \chi_2,\Tilde{\Lambda}, e_0 )$ and calculate the statistical errors. 

For CE and ET, we use $f_{\rm low} = 5$Hz and $1$Hz, respectively, and the $f_{\rm up}$ is computed as the frequency corresponding to the innermost stable circular orbit (ISCO) for the remnant Kerr BH. The full expression for Kerr ISCO can be found in Appendix C of ~\cite{Favata:2021vhw}. For BNS, we restrict $f_{\rm up} =1500$ Hz for CE and ET. The noise PSD for CE is taken from Eq.~(3.7) of~\cite{Kastha:2018bcr}. For ET, the noise PSD is taken from \cite{Hild:2010id}. Additionally, to account for the triangular shape of ET, we include a factor of $\sqrt{3}/2$ in the amplitude of the waveform. 

\section{Error estimation of tidal deformability for Binary Neutron Star}\label{sec:results}
\begin{figure*}
    \centering
    \includegraphics[width=\linewidth]{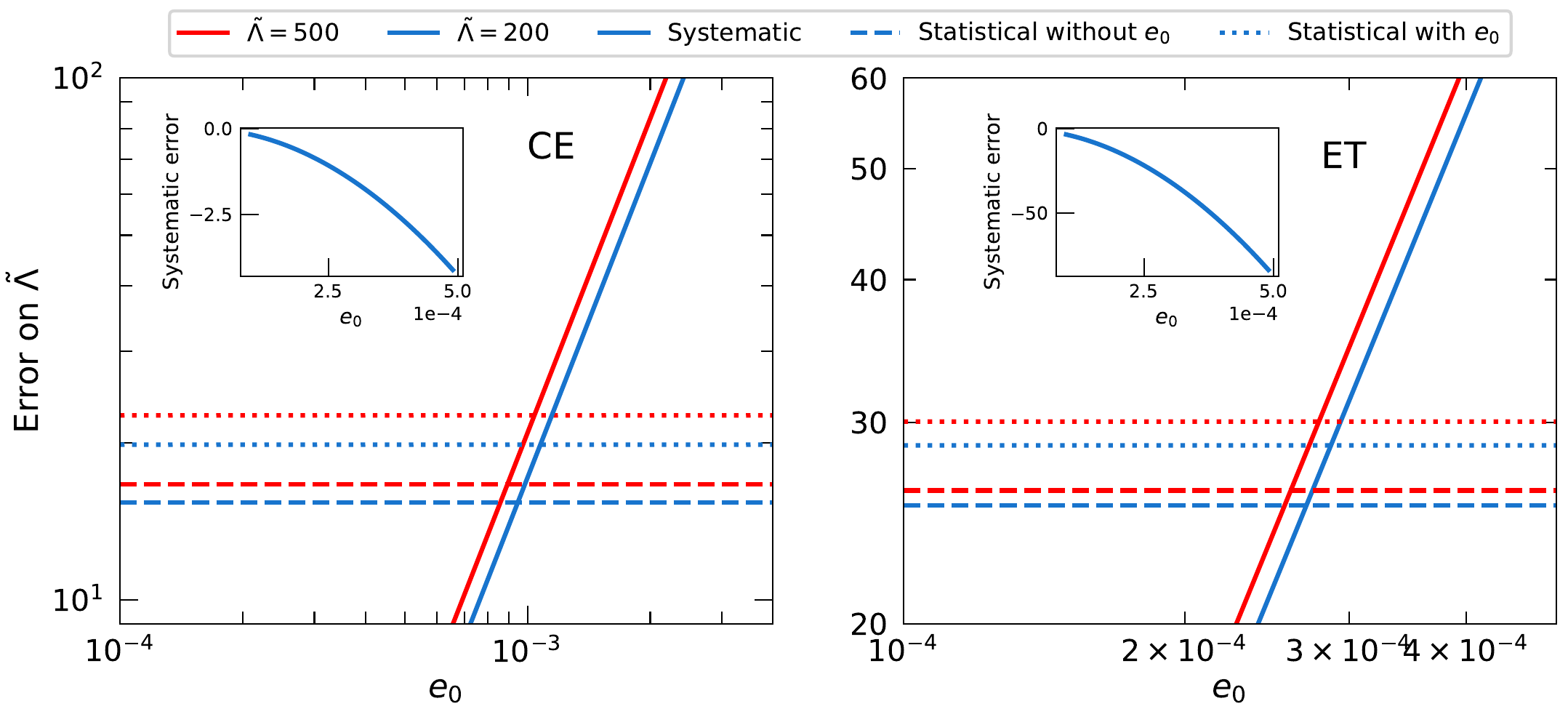}
    \caption{(Color online) Figure shows the systematic (solid lines) and statistical errors on the tidal deformability parameter $\Tilde{\Lambda}$ as a function of initial eccentricity $e_0$ (at a reference frequency of $10$Hz) for equal mass $(1.4 + 1.4)M_{\odot}$ BNS. The left panel is for CE and the right panel is for ET. The dashed (dotted) lines represent the statistical errors without (with) eccentricity included in the waveform model. The dimensionless spins are $\chi_{1,2} = 0.05$. The source is located at a luminosity distance of $100$ Mpc. The systematic errors exceed statistical errors at $e_0 \sim 10^{-3}$ ($\sim 3\times10^{-4}$) for CE (ET). The inset Figures show the sign of the systematic error on a linear scale for $\Tilde{\Lambda} =200$. The inclusion of eccentricity increases the statistical error by a factor of $\lesssim 2$ for both CE and ET.}
    \label{fig:BNS}
\end{figure*}

We compare the statistical and systematic error (due to eccentricity) on $\Tilde{\Lambda}$ 
for an equal mass BNS system $(m_1, m_2)=(1.4, 1.4) M_{\odot}$ located at a distance of $100$ Mpc ($z=0.022255$). The spins for both NSs are assumed to be aligned with the orbital angular momentum having dimensionless spin magnitudes $\chi_1=\chi_2=0.05$. We choose two values of tidal deformability parameter $\Tilde{\Lambda} = 200$ and $500$. The system has SNR $\sim 508$ and $\sim 181$ in CE and ET, respectively. It is possible to measure smaller eccentricity with the 3G detectors owing to their improved sensitivity, especially at lower frequencies. CE and ET, at their design sensitivity, can resolve the typical BBH's eccentricity $e_0\sim 5\times 10^{-3}$ and $\sim 10^{-3}$ at $10$Hz GW frequency, respectively~\cite{Lower:2018seu,Saini:2023wdk}. For BNS, even smaller eccentricity can be measured since it spends more GW cycles in the detector band. 

Figure~\ref{fig:BNS} shows the systematic and statistical errors in $\Tilde{\Lambda}$ for CE (left) and ET (right). Solid (slanted) lines show systematic errors. The systematic errors are compared with the statistical errors without including eccentricity in the parameter space (shown in dashed lines). As the eccentricity increases, the systematic errors in $\Tilde{\Lambda}$ increase and become comparable to the statistical errors. For CE, the systematic error on $\Tilde{\Lambda}$ crosses the statistical errors at $e_0 \sim 10^{-3}$. The systematic biases become more severe for ET where the cross-over happens at even smaller eccentricity $e_0\sim 3 \times 10^{-4}$. Since the source enters earlier in the ET band (at $1$Hz) compared to CE (at $5$Hz), it spends more time in the ET band. Moreover, the source had a relatively larger eccentricity when it entered the ET band. Therefore systematic bias becomes dominant at a smaller eccentricity for ET. Note that for $\Tilde{\Lambda}=500$, systematic errors exceed statistical errors at a slightly smaller $e_0$ compared to $\Tilde{\Lambda}=200$. The inset plots in Figure~\ref{fig:BNS} show the sign of the systematic errors on $\Tilde{\Lambda}$ on the linear scale.

We also quantify the impact of including eccentricity in the waveform model on the statistical errors for $\Tilde{\Lambda}$. These are denoted by the dotted lines in Figure~\ref{fig:BNS}. The statistical errors for $\Tilde{\Lambda}$ increase slightly when eccentricity is incorporated in the waveform model. Statistical errors on $\Tilde{\Lambda}$ increase by a factor of $\lesssim 2$ for both CE and ET. This is expected because measuring eccentricity along with other binary parameters spreads the information present in the signal among estimated parameters. The statistical error on  $\Tilde{\Lambda}$, when eccentricity is included in the waveform, has little dependence on the value of $e_0$ and appears almost straight line in the figure.

\subsection{Implications of biased \texorpdfstring{$\Tilde{\Lambda}$}{$  ̃Λ$} for NS EoS}
For a given EoS, the dimensionless tidal deformability $\hat{\lambda}$ for a NS can be expressed as a function of the source frame NS mass $(m_{\rm s})$ and can be obtained by solving the {\it Tolman-Oppenheimer-Volkoff} equations~\cite{PhysRev.55.364, PhysRev.55.374}. For our purpose, we use the polynomial fit for $\hat{\lambda}$ in terms of $m_{\rm s}$ given in Eq.(2.9) of \cite{Dhani:2022ulg}. This fit assumes the MPA1 EoS and is given as
\begin{align}\label{eq:eos}
    \text{log}_{10}\hat{\lambda} = \mathcal{F}(m_s) 
 \nonumber  &=  -1.21 m_s^4 + 7.80 m_s^3 -18.2 m_s^2  \\ &+ 16.5 m_s - 1.46 \,.
\end{align}
For equal mass NSs, the individual tidal deformability of a NS is equal to the reduced tidal deformability i.e., $\Tilde{\Lambda} = \hat{\lambda}$. Note that there are several EoS for NS that have not been ruled out. We take a sample case with the assumption of MPA1 being the true EoS. The calculation is simplified due to the analytical relation of Eq.~\eqref{eq:eos} and serves to show our result regarding the deviation of the inferred EoS from the true one. 

For NS with mass $m_s=1.4 M_{\odot}$, Eq.~\eqref{eq:eos} predicts $\Tilde{\Lambda} = 528.28$ which represents the true value of $\Tilde{\Lambda}$. Systematic bias on $\Tilde{\Lambda}$ and $m_s$ introduces a systematic shift in their inferred value, i.e., $\Tilde{\Lambda}+\Delta\Tilde{\Lambda}$ and $m_s+\Delta m_s$. The shift in the value of $\Tilde{\Lambda}$ and $m_s$ can either be positive or negative depending on the sign of the systematic bias, which in turn depends on the correlation of $\Tilde{\Lambda}$ and $m_s$ with $e_0$. The inset plots in Figure~\ref{fig:BNS} show that the systematic errors on $\Tilde{\Lambda}$ are negative. In other words, neglecting eccentricity \emph{underestimates} the value of $\Tilde{\Lambda}$. This can be qualitatively understood by looking at the number of GW cycles due to eccentricity and tidal deformability. Increasing the magnitude of the tidal deformability parameter leads to an increase in the number of inspiral cycles, while the number of cycles decreases with an increase in eccentricity~\cite{Favata:2013rwa}. An eccentric GW signal has less number of cycles compared to a circular signal, which is also the case for lower tidal deformability. Thus, when a quasicircular template is used for analyzing an eccentric signal, it underestimates the value of tidal deformability.

The source frame mass $m_s$ is biased through the detector frame mass $m_{\rm det}$ via the relation $m_s = m_d/(1+z)$. The systematic errors on $m_s$ can be calculated as~\cite{Bhat:2022amc}   
\begin{equation}
    \Delta m_s = \bigg(\frac{\partial m_s}{\partial m_{\rm det}}\bigg)\Delta m_{\rm det} + \bigg(\frac{\partial m_s}{\partial z}\bigg) \Delta z \,.
    \label{eq:ms_sys}
\end{equation}
We assume that there are no systematic errors on $z$ ($\Delta z =0$). We also assume that the systematic error is the same on both the NS in the binary i.e., $\Delta m_{\rm det} =(1/2) \Delta M$. The systematic errors in $M=\mathcal{M} \eta^{-3/5}$ can be computed as,
\begin{eqnarray} \label{eq:sys_M}
      \Delta M = \Big(\frac{\partial M}{\partial \mathcal{M}} \Big) \Delta \mathcal{M} + \Big( \frac{\partial M}{\partial \eta}\Big) \Delta \eta \,.
\end{eqnarray}
Substituting the derivatives and simplifying we obtain
\begin{equation}\label{eq:mdet_sys}
    \Delta m_{\rm det} = \frac{M}{2} \Big[\Delta (\ln \mathcal{M}) - \frac{3}{5} \Delta (\ln \eta) \Big] \,.
\end{equation}
The systematic errors $\Delta(\ln\mathcal{M})$ and $\Delta(\ln\eta)$ are computed using the Cutler-Vallisneri formalism. Substituting $\Delta m_{\rm det}$ in Eq.~\eqref{eq:ms_sys}, we obtain $\Delta m_s$. The inferred values of $(\Tilde{\Lambda}, m_s)$ are determined by adding the systematic error to the true values of $\Tilde{\Lambda}$ and $m_s$.

\begin{figure}
    \centering
    \includegraphics[width=\linewidth]{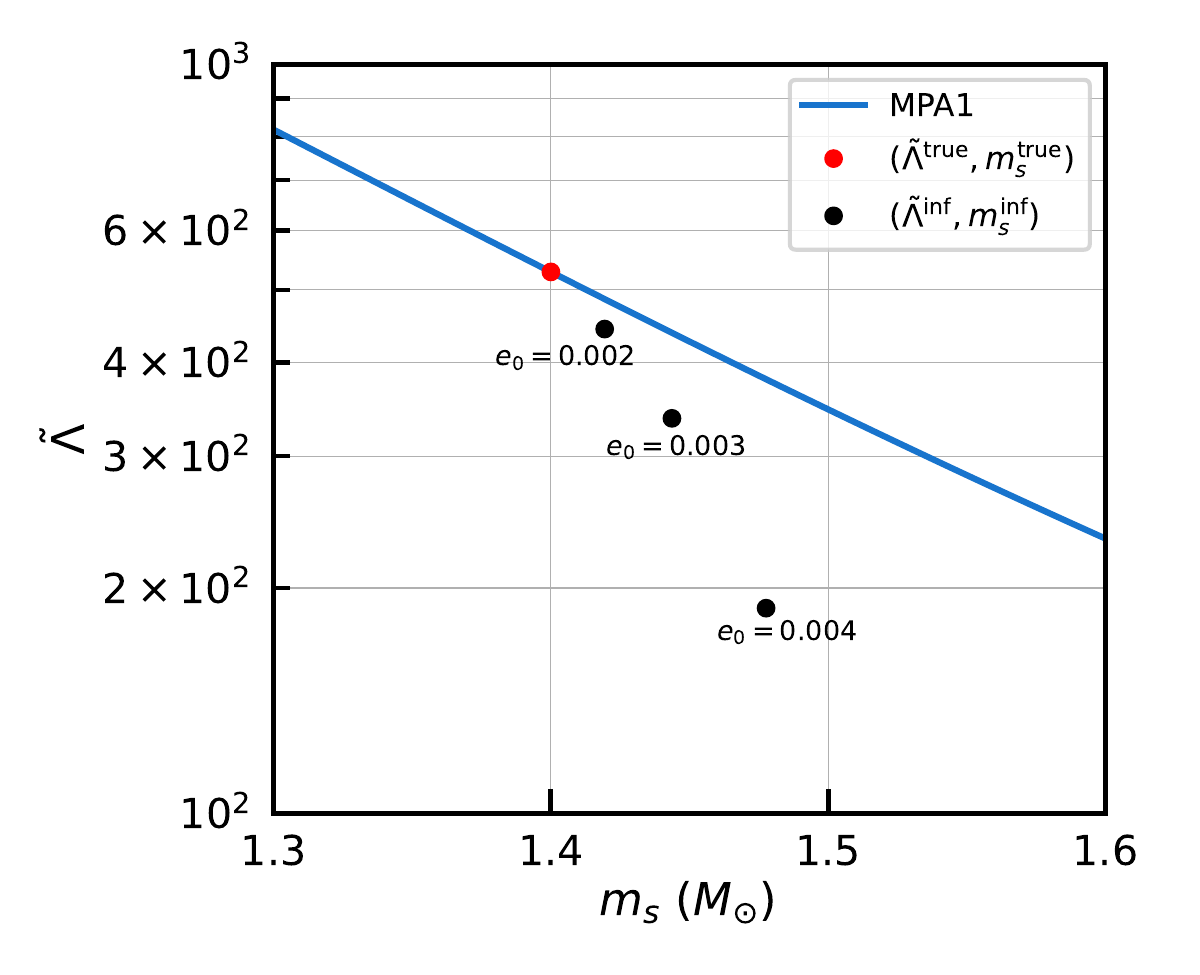}
    \caption{(Color online) The blue curve shows the MPA1 EoS (with $\hat{\lambda} = \Tilde{\Lambda})$ in Eq.~\eqref{eq:eos}. The red dot represents the true value of $\Tilde{\Lambda}$ and $m_s$. Systematic biases in $\Tilde{\Lambda}$ and $m_s$ shift their inferred values $(\Tilde{\Lambda}^{\rm inf}, m_s^{\rm inf})$ from the true values $(\Tilde{\Lambda}^{\rm true}, m_s^{\rm true})$. The black dots denote $(\Tilde{\Lambda}^{\rm inf}, m_s^{\rm inf})$ for different initial eccentricity $e_0$ defined at $10$Hz GW frequency. We consider an equal mass BNS $(1.4+1.4)M_{\odot}$ with $\chi_{1,2} =0.05$ at $100$ Mpc in CE.}
    \label{fig:eos_bias}
\end{figure}

Figure~\ref{fig:eos_bias} shows the shift in the inferred values of $\Tilde{\Lambda}$ and $m_s$ for different eccentricity values. The red dot denotes the true value of $\Tilde{\Lambda}$ and $m_s$. As the value of eccentricity increases, the inferred values of $\Tilde{\Lambda}$ and $m_s$ (shown by black dots) depart significantly from their true value, biasing the EoS inference. Note that the statistical errors in $\Tilde{\Lambda}$ and $m_s$ are small, thus we do not plot the statistical error ellipse around the dots. At $e_0\sim 0.003$, the inferred values $(\tilde{\Lambda}^{\rm inf}, m_{\rm s}^{\rm inf})$ differ significantly from the true values $(\tilde{\Lambda}^{\rm true}, m_{\rm s}^{\rm true})$.

\subsection{Implications of biased \texorpdfstring{$\Tilde{\Lambda}$}{$  ̃Λ $} for redshift inference via Love siren method} \label{sec:Love siren}
This section analyzes the consequences of the biased $\Tilde{\Lambda}$ estimate on the redshift inference via the Love siren method \cite{Messenger:2011gi,DelPozzo:2015bna,Chatterjee:2021xrm,Dhani:2022ulg,PhysRevD.106.123529}. Unlike the previous analysis, where errors on redshift were neglected, we now calculate the errors on redshift measurement using the Love siren technique. The observations of GW signals provide the detector frame mass $m_{\rm det}$, which is related to the source frame mass $m_s$ by the cosmological redshift 
\begin{equation}\label{eq:mz}
    m_{\rm {det}} = m_{s} (1+z) \,.
\end{equation}
For a given EoS, tidal deformability can be expressed entirely in terms of $m_s$. Therefore, the measurement of tidal deformability can be used to infer $m_s$, which then can be used to infer the redshift $z$. A biased tidal deformability will result in a biased value of $m_{s}$, leading to a biased $z$ measurement. Note that the accuracy with which the redshift can be estimated from the tidal effects would depend upon the EoS and the redshift~\cite{Messenger:2011gi}.
Once again we consider a $(1.4, 1.4) M_{\odot}$ BNS at $100$ Mpc $(z= 0.022255)$ with $\chi_{1,2}=0.05$. From Eq.~\eqref{eq:eos}, $\hat{\lambda} \equiv \Tilde{\Lambda} = 528.28$. 

The inferred redshift will be biased through $m_s$ due to systematic bias in $\Tilde{\Lambda}$ according to Eq.~\eqref{eq:eos} and through systematic bias in $m_{\rm det}$. Using Eq.~\eqref{eq:mz}, the systematic error on $z$ can be written as
\begin{eqnarray}
    \Delta z = \Big(\frac{\partial z}{\partial m_{\rm{det}}} \Big) \Delta m_{\rm{det}} + \Big( \frac{\partial z}{\partial m_{s}} \Big) \Delta m_{s} \,.
\end{eqnarray}
To quantify the individual effect of systematic errors due to $m_{\rm det}$ and $m_s$, we consider three scenarios; (a) when $\Delta m_{\rm det} = 0$ i.e., true detector frame mass is known, (b) when $\Delta m_{s}= 0$ i.e., true source frame mass is known and, (c) when both systematic biases on $m_{\rm det}$ and $m_s$ are included. The $\Delta m_{\rm{det}}$ is calculated from Eq.~\eqref{eq:mdet_sys} and $\Delta m_s$ is obtained from Eq.~\eqref{eq:eos}
\begin{equation}
   \Delta m_s = \bigg(\frac{\partial m_s}{\partial \Tilde{\Lambda}} \bigg) \Delta \Tilde{\Lambda} = \bigg(\frac{1}{2.30259 \Tilde{\Lambda} \, \mathcal{F}' }\bigg) \Delta \Tilde{\Lambda} \,,
\end{equation}
where $\mathcal{F}' = \partial{\mathcal{F}}/\partial{m_s}$. We compare the systematic errors with the $1\sigma$ statistical errors. The statistical errors on redshift $(\delta z)$ are computed via standard statistical error propagation
\begin{align}
    \Sigma_{zz} = \delta z^2  & = \Big(\frac{\partial z}{\partial m_s} \Big)^2 \Sigma_{m_sm_s} +  \Big(\frac{\partial z}{\partial m_{\rm det}} \Big)^2 \Sigma_{m_{\rm det} m_{\rm det}}  \nonumber\\
    &+2 \Big(\frac{\partial z}{\partial m_{\rm det}} \Big)\Big(\frac{\partial z}{\partial m_s} \Big) \Sigma_{m_s m_{\rm det}} \,.
\end{align}
The covariances $\Sigma_{m_s m_s},\, \Sigma_{m_{\rm det}m_{\rm det}}$, and $\Sigma_{m_s m_{\rm det}}$ are given by the following relations
\begin{subequations}
\begin{align}
  \Sigma_{m_sm_s} &= \Big(\frac{\partial m_s}{\partial \Tilde{\Lambda}} \Big)^2 \Sigma_{\Tilde{\Lambda}\Tilde{\Lambda}}\,,\\
\Sigma_{m_{\rm det}m_{\rm det}} &= \Big(\frac{\partial m_{\rm det}}{\partial (\ln\mathcal{M})}\Big)^2 \Sigma_{\ln\mathcal{M}\ln\mathcal{M}} \\ \nonumber 
     &+\Big(\frac{\partial m_{\rm det}}{\partial (\ln\eta)}\Big)^2 \Sigma_{\ln\eta\ln\eta}  \\ \nonumber 
    &+2 \Big(\frac{\partial m_{\rm det}}{\partial (\ln\mathcal{M})}\Big)  \Big(\frac{\partial m_{\rm det}}{\partial (\ln\eta)}\Big) \Sigma_{\ln\mathcal{M}\ln\eta} \,, \\     
  \Sigma_{m_sm_{\rm det}} &= \Big(\frac{\partial m_s}{\partial \Tilde{\Lambda}} \Big)\Big(\frac{\partial m_{\rm det}}{\partial (\ln\mathcal{M})}\Big) \Sigma_{\Tilde{\Lambda}\ln\mathcal{M}}\\ \nonumber &+\Big(\frac{\partial m_s}{\partial \Tilde{\Lambda}} \Big)\Big(\frac{\partial m_{\rm det}}{\partial (\ln\eta)}\Big) \Sigma_{\Tilde{\Lambda}\, \ln\eta} \,.
\end{align}
\end{subequations}  
The covariance terms $(\Sigma_{\Tilde{\Lambda} \Tilde{\Lambda}}, \cdots)$ can be read off from the covariance matrix [described in Eq.~\eqref{eq:covar}].

\begin{figure*}
    \centering
    \includegraphics[width=\linewidth]{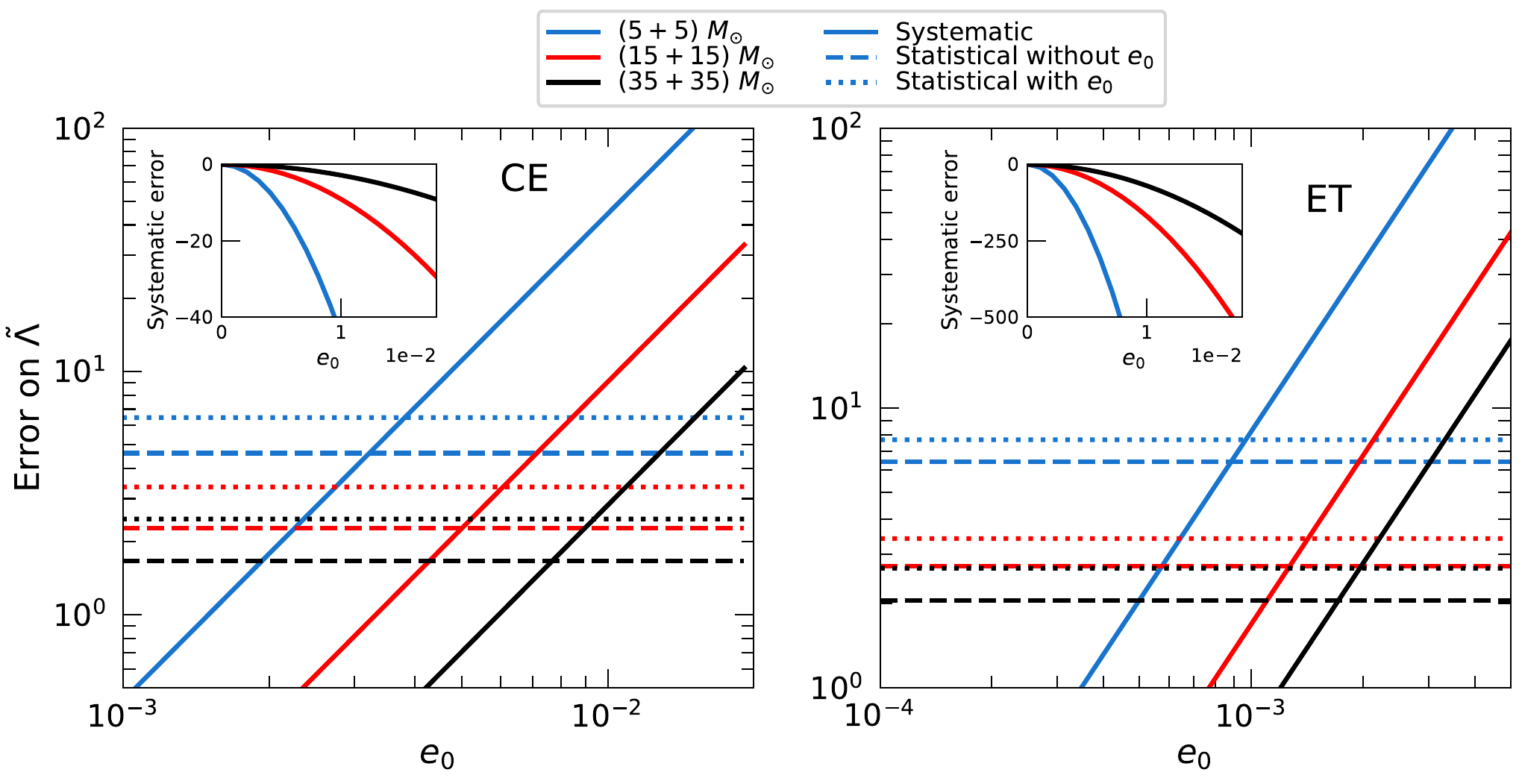}
    \caption{(Color online) Systematic (solid lines) and statistical errors on the tidal deformability parameter $\Tilde{\Lambda}$ as a function of eccentricity $e_0$ (at a reference frequency of $10$ Hz) for CE (left) and ET (right). Different colors show binaries of different masses. We consider equal mass BBHs with $\chi_{1,2} = 0.2$ at $500$ Mpc. The dashed (dotted) lines represent the statistical errors without (with) eccentricity included in the waveform model. For $(5+5)M_{\odot}$, the systematic error crosses over statistical error at $e_0 \sim 3 \times 10^{-3}$ ($ \sim 8 \times 10^{-4}$) for CE (ET).  The inset figures show the sign of systematic errors in linear scale. The inclusion of eccentricity increases the statistical error by a factor of $\lesssim 2$.} 
    \label{fig:BBH}
\end{figure*}

Figure~\ref{fig:z_bias} shows the biased redshift $z_b$ due to $m_s$ and $m_{\rm det}$. The black horizontal line represents the true value of the redshift. Systematic bias increases as the value of $e_0$ increases. The biases due to $m_s$ and $m_{\rm det}$ are of negative and positive sign, respectively. The bias due to $m_s$ increases more rapidly compared to $m_{\rm det}$ bias. Therefore, the combined systematic bias is negative. This means that an unaccounted eccentricity will lead to the underestimation of the redshift of the GW event. The shaded region in Figure~\ref{fig:z_bias} shows the $1\sigma$ statistical errors around the true value of $z$. The combined systematic bias becomes comparable to the statistical errors at $e_0\sim 10^{-3}$. Therefore, if a BNS system with a small eccentricity $(e_0 \sim 10^{-3})$ is detected in 3G detectors, the unmodeled eccentricity will significantly bias the redshift inference. 

\section{Error estimation of tidal deformability for binary black hole}\label{sec:BBH nature}

In this section, we study the effect of neglecting eccentricity on the test of Kerr nature of BH. The tidal deformability is predicted to be zero for black holes in GR~\cite{Pani:2015hfa,Landry:2015zfa,PhysRevLett.114.151102,Chakrabarti:2013lua,LeTiec:2020spy}. On the other hand, exotic compact objects can have non-zero tidal deformability~\cite{Uchikata:2016qku,Cardoso:2017cfl,Mendes:2016vdr}. The measurement of tidal deformability for BHs can be considered a null test of the Kerr nature of BH, the biased value of $\Tilde{\Lambda}$ will indicate a false detection of a non-BBH.

We consider three representative equal mass BBH systems with total mass $M=(10 M_{\odot},30 M_{\odot},70 M_{\odot})$. The dimensionless spin magnitudes are assumed to be $\chi_{1,2}=0.2$. The luminosity distance to each source is fixed at $500$Mpc. These systems $(10 M_{\odot},30 M_{\odot},70 M_{\odot})$ have SNR $\sim (313,778,1550)$ and $\sim (112,276,535)$ in CE and ET, respectively. For all BBH systems, we obtain statistical and systematic errors around $\Tilde{\Lambda} = 0$. 

Figure~\ref{fig:BBH} shows the statistical (dashed line) and systematic error (solid line) on $\Tilde{\Lambda}$ as a function of $e_0$ for CE (left) and ET (right). Systematic errors increase as $e_0$ increases and become comparable to the statistical errors. For lower mass systems, the systematic error becomes dominant at a lower value of $e_0$. Note that the statistical errors on $\Tilde{\Lambda}$ are smaller for higher mass systems since these systems have larger SNR. Though statistical errors are similar for CE and ET, systematic errors are larger for ET. This is due to the longer inspiral of binary in the ET band. Since the binary enters the ET band at $1$Hz and the CE band at $5$Hz, it spends more GW cycles in the ET band. Moreover, the binary had a larger eccentricity at $1$Hz when it entered the ET band. Therefore, the systematic errors dominate the statistical errors at a smaller eccentricity for ET. For $10 M_{\odot}$ in CE, systematic bias crosses the statistical error at $e_0\sim 3\times 10^{-3}$ while for ET the cross-over happens at $e_0 \sim 8 \times 10^{-4}$. 

The dotted, horizontal lines in Fig.~\ref{fig:BBH} show the statistical errors on  $\Tilde{\Lambda}$ when eccentricity is included in the waveform. This leads to a slight increase in their magnitude by a factor of $\lesssim 2$. The statistical error on  $\Tilde{\Lambda}$, when eccentricity is included in the waveform, shows little dependence on the value of $e_0$ and appears almost as a straight line in the figure.

The systematic bias can shift the value of $\Tilde{\Lambda}$ to the positive or negative side depending upon the sign of the systematic bias, which depends on the correlation of $\Tilde{\Lambda}$ with $e_0$. The inset plots in Fig.~\ref{fig:BBH} shows the sign of the systematic bias in $\Tilde{\Lambda}$ on a linear scale. Systematic bias on $\Tilde{\Lambda}$ is negative, meaning the neglect of eccentricity will mimic BBHs with those classes of compact objects that have negative $\Tilde{\Lambda}$. Theoretically proposed exotic compact objects like gravastars~\cite{Mazur:2004fk} are known to exist with negative tidal deformability~\cite{Uchikata:2016qku,Mottola:2023jxl}. 

\section{Conclusions} \label{sec:conclusion}
Third-generation (3G) GW detectors are expected to observe a large number of BNSs along with BBHs. A fraction of BNSs, depending on their formation scenarios, can possess residual eccentricity while entering the frequency band of 3G detectors: Cosmic Explorer and Einstein Telescope. We studied the impact of unmodeled eccentricity on the measurement of tidal deformability. Since GW observations will yield extremely precise measurements of tidal deformability, we find that even very small eccentricity leads to the biased measurement of tidal deformability parameter $\Tilde{\Lambda}$. Since eccentricity and tidal deformability affect the GW waveform at relatively lower and higher post-Newtonian order, respectively, their measurement is naively expected to have negligible effects on each other. However, our study indicates that the precise measurement of the tidal deformability in 3G detectors can be biased even at a very small eccentricity. This is due to the combined effect of overall improved sensitivity of 3G detectors, longer GW signal in the detector band due to enhanced sensitivity at lower frequencies, and the best sensitivity at higher frequencies where tidal effects start dominating. 

Since a BNS system spends more GW cycles in the frequency band of ET compared to CE, systematic bias dominates the statistical errors at a smaller eccentricity for ET. Systematic errors on $\Tilde{\Lambda}$ become greater than statistical errors at $e_0\sim 3\times 10^{-4}$ (defined at $10$Hz GW frequency) in ET, while for CE systematic errors cross statistical errors at $e_0\sim 10^{-3}$. Considering MPA1 as the true EoS for NS, we find that systematic errors in $\Tilde{\Lambda}$ significantly bias the EoS inference at $e_0\sim 3\times10^{-3}$. 

Additionally, we studied the bias introduced in redshift estimation through the `Love siren' method. Systematic errors in $\Tilde{\Lambda}$ propagate to the source frame NS mass. The systematic errors on $\Tilde{\Lambda}$ (and hence on source frame mass) and detector frame mass significantly bias the redshift measurement leading to its underestimation. The combined systematic errors in redshift become comparable to the statistical errors at $e_0\sim 10^{-3}$. Finally, we show that the biased $\Tilde{\Lambda}$ for BBHs, indicates a deviation from the Kerr nature. 

We also studied the impact of including eccentricity in the waveform on $\Tilde{\Lambda}$ measurement. The measurement of eccentricity along with other binary parameters is found to have a mild effect, increasing the statistical errors on $\Tilde{\Lambda}$ by a factor of $\lesssim 2$. 

Since 3G detectors will be observing a large number of GW signals from merging compact binaries, leading to more precise measurements of binary parameters by combining information from a large GW catalog. The statistical errors decrease as $\sim 1/\sqrt{N}$, where $N$ is the number of events. Thus, systematic errors can become severe even at a smaller value of eccentricity. In conclusion, our study highlights the need for incorporating eccentricity in GW waveforms.  

Note that our waveform model describes only the inspiral of a binary system and neglects the merger-ringdown part. Moreover, the spins are assumed to be aligned with the orbital angular momentum (non-precessing). For BNS, the merger-ringdown and spins are expected to be small. Hence, these assumptions are likely to have negligible effect on our results. However, including these effects in the waveform model would only improve the measurement of tidal effects~\cite{Lackey:2013axa,Messenger:2013fya}, making the systematic errors an even more severe issue. \\

\section*{Acknowledgments}
We thank K. G. Arun for the encouragement to pursue this problem, discussions, and useful comments on the manuscript. The authors also thank Anuradha Gupta, B. S. Sathyaprakash, Nathan Johnson-McDaniel, Rahul Kashyap, Archisman Ghosh, Arnab Dhani, Parthapratim Mahapatra, and Sajad Ahmad Bhat for valuable discussions. P. S. acknowledges support from the Department of Science and Technology and the Science and Engineering Research Board (SERB) of India via Swarnajayanti Fellowship Grant No.~DST/SJF/PSA-01/2017-18. P. D. R. and P. S. acknowledge the support from the Infosys Foundation. 
\bibliographystyle{apsrev}
\bibliography{refs}

\end{document}